\documentclass{article}

\usepackage[preprint]{nips_2018}

\usepackage[utf8]{inputenc} 
\usepackage[T1]{fontenc}    
\usepackage{hyperref}       
\usepackage{url}            
\usepackage{booktabs}       
\usepackage{amsfonts}       
\usepackage{nicefrac}       
\usepackage{microtype}      
\usepackage{textcomp}
\usepackage{graphicx}

\title{MRAM Co-designed Processing-in-Memory CNN Accelerator for Mobile and IoT Applications}

\author{
  Baohua Sun, Daniel Liu, Leo Yu, Jay Li, Helen Liu, Wenhan Zhang, Terry Torng\\
  Gyrfalcon Technology Inc.\\
  1900 McCarthy Blvd. Suite 208, Milpitas, CA 95132 \\
  \texttt{\{baohua.sun, daniel.liu, wenhan.zhang, terry.torng\}@gyrfalcontech.com} \\
}

\begin{document}

\maketitle

\begin{abstract}
  We designed a device for Convolution Neural Network
  applications with non-volatile MRAM memory and
  computing-in-memory co-designed architecture. It has been successfully
  fabricated using 22nm technology node CMOS Si process.
  More than 40MB MRAM density with 9.9TOPS/W are provided. 
  It enables multiple models within one single chip for mobile and IoT device applications.
\end{abstract}

\section{Introduction}

Artificial Intelligence (AI) is recognized as one of the key
technology in the Fourth Industrial Revolution. AI also known
as machine learning has been around for a long while [7] only
recently becoming more advanced, popular and mature. Deep
learning has found its applications in a wide variety of tasks
such as computer vision, image and speech recognition, machine
translation, robotics, and medical image processing, etc [1].
Hardware acceleration of deep learning tasks come in perfect
timing to offer a multi-fold speed improved and functional logic
processor with dedicated Application-Specific Integrated
Circuit (ASIC) and its memory storage system.

Convolutional Neural Network (CNN) models are
successful in computer vision tasks [6][8]. It repeatedly
executes convolution operations on the input image, thus
power-efficient ASICs are highly desirable for IoT
implementations. Besides that, the CNN model sizes are big,
which requires large size of memories. For computing-in-memory
architectures, SRAM solution which provides low
memory density becomes the bottleneck. Sun, et al. (2018) has designed a
Convolutional Neural Networks Domain Specific Architecture
(CNN-DSA) accelerator for extracting features out of an input
image [2]. It processes 224x224 RGB images at 140fps with
ultra-power-efficiency, a record of 9.3 TOPS/Watt and peak power less
than 300mW. This architecture mainly focuses on inference,
rather than training. Its CNN-DSA hardware system based on
28nm node is produced successfully with all internal SRAM
memory only. Its APiM (AI Processing in Memory)
architecture can save power and increase speed, but it requires
large memory size with better power efficiency control. 

Magneto-resistive Random Access Memory (MRAM) is a
high speed Non-Volatile Memory (NVM) that can provide
unique solutions which improve overall system performance in
a variety of areas including data storage, industrial controls,
networking, and others. The recent development of magnetic 
tunnel junction (MTJ) – a key element of the MRAM device,
enables fundamentally a fast, reliable read and write operation
MRAM circuit. Spin Torque Transfer (STT) MRAM is an
emerging memory which possesses an excellent combination of
density, speed, and non-volatility [3]. STT MRAM is
considered as the most promising NVM when compared to
existing SRAM, eFlash, and ReRAM in terms of energy
efficiency, endurance, speed, extendibility and scaling [4].

Thus, APiM using MRAM memory would provide better
solution for larger memory size and better power efficiency of
enabling large CNN model or multiple CNN models. We introduce our unique
computational architecture by implementing the CNN in a
matrix form, with each element consisting of advanced memory
such as a non-volatile memory device. With MRAM co-designed,
large sized models could be possibly loaded into CNN chips.
This can be used in mobile, IoT
and embedded smart device systems in real world.

\section{STT MRAM technology}
\begin{figure}
  \centering
  \includegraphics[width=0.35\linewidth]{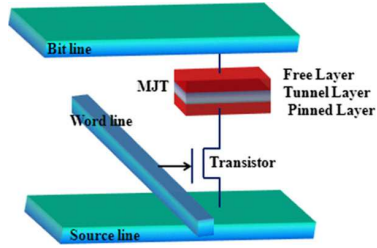}
  \caption{STT MRAM cell image.}
  \label{Figure_1}
\end{figure}

We successfully implemented advanced technology node of 22nm on CMOS, SRAM and
emerging STT MRAM. More than 40MB MRAM density was
embedded into our CNN engines. This is 4.5x increase of
memory compared to Sun, et al. (2018) SRAM based CNN-DSA, which
total memory size is about 9MB as a record [2]. Figure \ref{Figure_1} showed a schematic
drawing of a STT MRAM memory cell image. MRAM only
requires few additional masks to be processed in between
BEOL metal layers as magnetic storage structures. 

MRAM leakage was believed to be low [4]. To measure
power leakage on our real chip, standby current was tested by
placing chip on the socket of the test board and power supply
set to working voltage. We compared room temperature vs 70
\textdegree{}C on MRAM vs SRAM leakage behavior. Voltage VDD 0.9V
and VDIO 2V were used. Table \ref{sample-table} shows the dynamic and standby
power comparing MRAM vs SRAM. It demonstrated
that MRAM leakage power was low at 5.5mW at room temp
and 7.2mW at 70\textdegree{}C. STT MRAM indeed
offers embedded 4-5X higher density of memory but with much
lower leakage power consumption.

\begin{table}
  \caption{MRAM vs SRAM dynamic and standby power at room and high Temperature(70\textdegree{}C)}
  \label{sample-table}
  \centering
  \begin{tabular}{cccc}
    \toprule
    \multicolumn{4}{c}{Power(mW)}                   \\
    \midrule
    Conditions     &       & MRAM & SRAM \\
    \midrule
    Room Tempreature & Dynamic  & $38.3$ & $39.2$     \\
    Room Tempreature & Standby  & $5.5$ & $34.3$     \\
    High Tempreature & Dynamic  & $35.4$ & $43.1$     \\
    High Tempreature & Standby  & $7.2$ & $136$     \\
    \bottomrule
  \end{tabular}
\end{table}

\section{MRAM based convolutional neural network accelerator architecture}
\label{gen_inst}

\subsection{CNN Matrix Processing Engine (MPE)}
The CNN algorithm is constructed by stacking multiple
computation layers for feature extraction and classification [5].
Modern CNNs achieve their superior accuracy by building a
very deep hierarchy of layers [6], which transform the input
image data into highly abstract representations called feature
maps (fmaps). The primary computation in the CNN layers is
performing convolutional operations. A layer applies filters on
the input fmaps to extract embedded characteristics and
generate the output fmaps by accumulating the weighted sums
(Wsums) and non-linear activations. 

\begin{figure}
  \centering
  \includegraphics[width=0.35\linewidth]{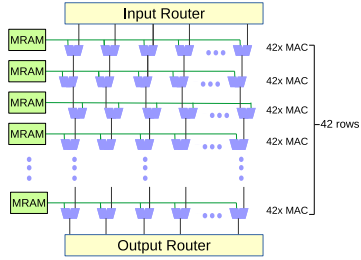}
  \caption{CNN Matrix Processing Engine (MPE).}
  \label{Figure_2}
\end{figure}

We designed a coprocessor for CNN acceleration. The CNN
processing block simultaneously performs 3x3 convolution on
2-D image at PxP pixel locations using input from input buffer
and filter coefficients from co-designed on-chip MRAM memory. Padding is applied
to PxP pixel locations for convolution. Each layer performs 3x3
convolution and bias can also be added. Then it is connected to
nonlinear activation operations. Max pooling operation follows
and shrink the output size by 4 times. Figure \ref{Figure_2} showed a CNN
Matrix Processing Engine implementation of M=14. 

Each CNN processing engine includes a CNN processing block, a first set
of memory buffers for storing imagery data and a second set of
memory buffers for storing filter coefficients. When two or
more CNN processing engines are configured on the IC
controller, the CNN engines connect to one another via a clockskew
circuit for cyclic data access and the number of I/O data
bus. Activations use 9 bits Domain Specific Floating Point
(DSFP), and model coefficients use 15 bits DSFP.

\subsection{System overview of STT MRAM based CNN accelerator}
\begin{figure}
  \centering
  \includegraphics[width=0.2\linewidth]{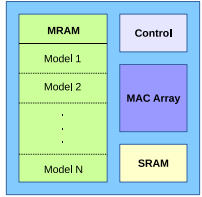}
  \caption{CNN block with memory array.}
  \label{Figure_3}
\end{figure}

CNN is implemented in hardware in the form that is very
similar to a memory array. This concept is a processor-in-memory
design architecture. Figure \ref{Figure_3} shows the block diagram for our chip architecture loading multiple models. 
It uses the CNN processing engine with co-designed MRAM architecture block, instead of using a regular SRAM.
Our chip architecture includes four major components, including on-chip MRAM, SRAM, MAC array and control unites. The MRAM loads the coefficients of multiple models in different locations of the memory. The SRAM stores the data inputs and intermediate results of activations, which is reused and read-write multiple times. The MAC array executes the convolution computation operations using coefficients in the MRAM and data activations in SRAM. And the control unit coordinates the above three components.

For a CNN based IC for artificial intelligence, data
must be provided as close to the CNN processing logic to save
power. In image processing, filter coefficients
and imagery data have different requirements for data access. Filter
coefficients need to be validly stored for long time, while the
imagery data are written and read more often. Since MRAM
has high endurance, preloaded models can be saved in MRAM
for some applications without requirement of external storage
and memory buffer to load model into chip. In view of this
requirement, STT MRAM is selected to be the NVM for storing
filter coefficients or weights. Such advanced memory has high
retention rate at 85 ℃ for 10 years which fulfills and allows the
purpose of memory needs for imbalanced read and write
operations.

MRAM allows bigger memory for loading models than SRAM. There are some key features on multi-task performance with multi-models capability in one single chip, such as voice and facial recognition simultaneously. Potential application can be identity verification from different aspect of biological characteristics, e.g. recognizing voice, face, fingerprints and gesture with four independent models in one single chip. In addition, MRAM AI chip also enables ensemble of multiple models in a single chip. 

\subsection{Inference with MRAM based CNN accelerator}
\begin{figure}
  \centering
  \includegraphics[width=0.3\linewidth]{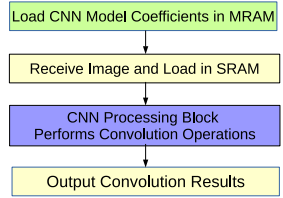}
  \caption{Inference workflow of MRAM CNN accelerator chip.}
  \label{Figure_4}
\end{figure}

Figure \ref{Figure_4} shows the working flow of using our AI MRAM chip. 
First, it loads the CNN coefficients in MRAM. Second, it
loads image into SRAM, which is fast for multiple read and write.
Third, coefficients and image data are sent to CNN processing
block for convolutional operations. At last, the convolution
results will be sent to host processor. Color coding
corresponds to the activated components in same color as in
Figure \ref{Figure_3}. 

Based on the measurement of the operating power, current, length
and background current while running the image classification,
the power consumption of coefficients memory is about $1/4$ of total power,
and rest part of chip consumes $3/4$ of total power. 
By fitting Ivdd (Vdd operation current) vs. frequency curve as SRAM chip, we got Ivdd is around three quarters of SRAM.
Based on the above measurements, the power efficiency of MRAM based CNN accelerator can be calculated as $9.3/[3/4+(1/4)*(3/4)] = 9.9$ TOPS/W. 
At 12.5Mhz, we processes 3x224x224 image at the speed of 35fps, which should be
sufficient for applications in IoT and mobile scenarios.

\section{Conclusion}
\label{conclu}
The 22nm device of STT MRAM memory co-designed with processing-in-memory CNN accelerator has been successfully fabricated. 
Compared with SRAM, the lower leakage of MRAM push the record of power efficiency to 9.9 TOPS/W with reliable read and write
operations. Image classification and voice recognition tasks were successfully executed simutaneously on one single chip. 
At 2nd Workshop on Machine Learning on the Phone and other Consumer Devices,
we will demo our MRAM CNN chip with multiple models on one single chip, and show the CNN chip with non-volatile feature. 
MRAM co-designed processing-in-memory CNN accelerator chip would be used for mobile, IoT, and smart device applications. 

\section*{References}
\medskip
\small
[1] I. Goodfellow, Y. Bengio, and A. Courville, "Deep Learning", the MIT
press, pp. 8-10, 2016.

[2] Baohua Sun, Lin Yang, Wenhan Zhang, “Ultra power-efficient CNN
domain specific accelerator with 9.3TOPS/W for mobile and embedded
applications”, CVPR 2018.

[3] A. D. Kent and D. C. Worledge (2015) Nature Nano. 10, 187. 

[4] Luc Thomas et al, “Basic principles, challenges, and opportunities of STT
MRAM for embedded memory application”, MSST 2017, (2017).

[5] Y. LeCun, K. Kavukcuoglu, and C. Farabet, “Convolutional networks
and applications in vision”, in Proc. IEEE Int. Symp. Circuits Syst.
(ISCAS), May/Jun. 2010, pp. 253–256.

[6] K. He, X. Zhang, S. Ren, and J. Sun, “Deep residual learning for image
recognition”, in Proc. IEEE Conf. CVPR, 2016.

[7] L.O. Chua and L. Yang, “Cellular neural networks: Applications”, IEEE
Transactions on Circuits and Systems, 35(10):1273-1290, 1988.

[8] K. Simonyan and A. Zisserman, “Very deep convolutional networks for
large-scale image recognition”, CoRR, abs/1409.1556, 2014. 

\end{document}